\documentclass[apj]{emulateapj}

\newcommand{\code}[1]{\texttt{#1}}
\newcommand*{\mesa}{\code{MESA}}
\newcommand*{\dStar}{\code{dStar}}
\newcommand*{\ecapture}{$e^{-}$-capture}
\newcommand*{\ecaptures}{$e^{-}$-captures}
\newcommand*{\bdecay}{$\beta^{-}$-decay}
\newcommand*{\bdecays}{$\beta^{-}$-decays}

\newcommand*{\cc}{$^{12}$C$+$$^{12}$C}

\newcommand*{\apjfigscale}{0.85}
\newlength{\apjcolwidth}
\newlength{\figwidth}
\newlength{\doublewide}
 \submitted{submitted to The Astrophysical Journal}
 
\begin{document}
 
 \title{Urca cooling pairs in the neutron star ocean and their effect on superbursts}
 
\author{Alex Deibel\altaffilmark{1,2}, 
Zach Meisel\altaffilmark{2,3},
Hendrik Schatz\altaffilmark{1,2,4},
Edward F. Brown\altaffilmark{1,2,4}, 
and Andrew Cumming\altaffilmark{2,5}}
\affil{
\altaffilmark{1}{Department of Physics and Astronomy, Michigan State University, East Lansing, MI 48824, USA; deibelal@msu.edu} \\
\altaffilmark{2}{The Joint Institute for Nuclear Astrophysics - Center for the Evolution of the Elements, East Lansing, MI 48824, USA} \\
\altaffilmark{3}{Department of Physics, University of Notre Dame,
Notre Dame, IN 46556, USA} \\
\altaffilmark{4}{National Superconducting Cyclotron Laboratory, Michigan State University,
East Lansing, MI 48824, USA} \\
\altaffilmark{5}{Department of Physics and McGill Space Institute, McGill University, 3600 rue University, Montreal, QC, H3A 2T8, Canada} \\
}
\shorttitle{URCA COOLING IN NEUTRON STAR SUPERBURSTS}
\shortauthors{DEIBEL ET AL.}

\begin{abstract}
An accretion outburst onto a neutron star deposits
hydrogen-rich and/or helium-rich material into the neutron star's
envelope. Thermonuclear burning of accreted material robustly
produces Urca pairs\textrm{---}pairs of nuclei that undergo cycles of $e^-$-capture and $\beta ^-$-decay. The strong $T^5$ dependence of the Urca cooling neutrino luminosity means that Urca pairs in the neutron star interior potentially remove heat from accretion-driven nuclear reactions. In this study, we identify Urca pairs in the neutron star's ocean \textrm{---} a plasma of ions and electrons overlaying the neutron star crust \textrm{---} and demonstrate that Urca cooling occurs at all depths in the ocean. We find that Urca pairs in the ocean and crust lower the ocean's steady-state temperature during an accretion outburst and that unstable carbon ignition, which is thought to trigger superbursts, occurs deeper than it would otherwise. Cooling superburst light curves, however, are only marginally impacted by cooling from Urca pairs because the superburst peak radiative luminosity $L_{\rm peak}$ is always much greater than the Urca pair neutrino luminosity $L_{\nu}$ in the hot post-superburst ocean.

\end{abstract}

 
\section{Introduction}

In a low-mass X-ray binary, an accretion outburst deposits hydrogen-rich and/or helium-rich material into the envelope of the neutron star primary. Nuclear burning of accreted material produces seed nuclei for the rapid proton-capture ($rp$) process and the $\alpha p$-process \citep{wallace1981}. The $rp$-process and $\alpha p$-process produce proton-rich nuclei with mass numbers in the range of $A\sim 60 \textrm{--} 100$ during unstable burning in Type I X-ray bursts \citep{schatz1998, woosley2004, cyburt2010} and superbursts \citep{schatz2003,keek2011,keek2012}. Ashes of nuclear burning are compressed deeper into the neutron star ocean during subsequent accretion outbursts.

Further accretion compresses envelope material into the neutron star's crust, where nonequilibrium nuclear reactions deposit $\approx 1\textrm{--}2 \, \mathrm{MeV}$ per accreted nucleon of heat \citep{haensel1990,gupta2007,haensel2008}. Crustal heating provides a heat flux into the overlying neutron star ocean, where the unstable ignition of $^{12}$C$+$$^{12}$C fusion triggers superbursts \citep{woosley1976, taam1978, brown1998,cumming2001,strohmayer2002}. Current superburst ignition models, however, require a larger heat flux than supplied by crustal heating to match observed superburst ignition depths \citep{cumming2001,cumming2006}. The source of extra heating is unknown, although some additional heat may be supplied by compositionally driven convection \citep{medin2011,medin2014,medin2015} and stable helium burning \citep{brown1998} in the neutron star envelope.

Nuclear burning ashes also contain pairs of nuclei that undergo
$e^{-}$-capture/$\beta^{-}$-decay cycles, known as Urca pairs, that cool the neutron star crust by neutrino emission \citep{bahcall1965,schatz2014}. Neutrino emission
from an Urca cycle scales as $T^5$ and potentially liberates a substantial amount of energy from the neutron
star's interior relative to the heat deposited by nonequilibrium reactions.  \citet{schatz2014} showed that for a large enough abundance of Urca pairs in the crust, Urca cooling could balance crustal heating when crust temperatures are $T \gtrsim 2 \times 10^{8} \, \mathrm{K}$. As a result, Urca cooling in the crust of hot neutron star transients, for example, MAXI~J0556-332 \citep{matsumura2011,homan2014}, impacts quiescent light curve predictions from crust thermal relaxation models \citep{deibel2015}.

Here we examine Urca pairs in the neutron star's ocean that potentially remove heat from the ocean near the depth of superburst ignition. Observed superbursts to date occur on accreting neutron stars with accretion rates in the range of $\dot{M} \approx (0.1 \textrm{--} 0.3) \, {\dot{M}_{\rm Edd}}$ \citep{wijnands2001_4U}, and superburst ignition depths are constrained to a range of $y_{\rm ign}\approx (0.5\textrm{--}3) \times 10^{12} \, \mathrm{g \ cm^{-2}}$ in these sources \citep{cumming2006}. We find that Urca pairs in the superburst ignition region are a factor of $\sim 100\textrm{--}1000$ less luminous than the pairs identified in the crust \citep{schatz2014}. As a result, the neutrino luminosity from ocean Urca pairs is always much less than the peak superburst luminosity ($L_{\nu} < L_{\rm peak}$), and ocean pairs have a minimal impact on cooling superburst light curve predictions. Neutrino cooling from crust Urca pairs, however, is sufficiently strong to limit the ocean's steady-state temperature to $T \lesssim  10^{9} \, \mathrm{K}$, and carbon ignition must occur at depths $y_{\rm ign} \gtrsim 7.8 \times 10^{10} \, \mathrm{g \ cm^{-2}}$.

In Section~\ref{s.pairs}, we calculate the neutrino luminosities from 85 Urca cycling pairs in the neutron star ocean and give a general formula for the neutrino luminosity associated with an Urca pair. In Section~\ref{s.superburst} we explore the impact of ocean and crust Urca pair neutrino cooling on carbon ignition depths and constraints on shallow heating provided by the shallow superburst ignition in 4U~1636--536. Superburst cooling light curves with Urca pairs in the ocean are calculated in Section~\ref{s.lightcurve}. We discuss our findings in Section~\ref{s.discussion}. 

\section{Ocean Urca pairs}\label{s.pairs}

When nuclear burning ashes are compressed into the neutron star ocean to mass densities $\rho \gtrsim 10^{6} \, \mathrm{g \ cm^{-3}}$, nuclei undergo nonequilibrium \ecaptures \ in the ambient degenerate electron gas \citep{bisnovatyi1979,sato1979}. At finite temperature, \ecaptures \ begin when $\mu_e \gtrsim |Q_{\rm EC}| \,-\, k_{\rm B}T $, where $Q_{\rm EC}$ is the \ecapture \ threshold energy and $\mu_e$ is the chemical potential of the electron gas. If the \ecapture \ daughter undergoes \bdecay \ before any other nuclear reactions,  the \ecapture \ parent and daughter form an \ecapture /\bdecay \ Urca cycle \citep{gamow1941}, and these nuclei form an Urca pair. Over an Urca cycle, a significant amount of energy is emitted as neutrinos, which effectively cools the neutron star interior \citep{gamow1941}.

\noindent 

Although the most abundant nuclear burning ashes in thermonuclear bursts are even-$A$ nuclei (e.g., Schatz et al. 1998), these nuclei do not typically form Urca pairs. For even-$A$ nuclei, with proton number $Z$ and neutron number $N$, $e^{-}$-captures typically proceed in two steps: (1) the even-$Z$
even-$N$ parent nucleus
$e^{-}$-captures into a low-lying state of the
odd-$Z$ odd-$N$ first \ecapture \ daughter, and then (2) the
odd-$Z$ odd-$N$ first \ecapture \ daughter immediately
$e^{-}$-captures into a low-lying state of the even-$Z$ even-$N$ second \ecapture \ daughter due to
odd--even mass staggering. Insufficient electron phase space exists
for the second $e^{-}$-capture daughter to undergo $\beta^{-}$-decay, and an \ecapture /\bdecay \ Urca cycle is therefore blocked. Under special circumstances even-$A$ nuclei may form an Urca cycle; however, this requires a specific set of excited-state energies and spin parities $J^{\pi}$ in the first and second $e^{-}$-capture daughters \citep{meisel2015}.

Nearly all odd-$A$ nuclei, with odd-$Z$ even-$N$ or even-$Z$ odd-$N$, may form an Urca cycle. For odd-$A$ nuclei, $|Q_{\rm EC}|$ increases monotonically for
more neutron-rich species. As a consequence, following an \ecapture \ onto an
odd-$A$ nucleus, a second \ecapture \ is energetically blocked. At finite temperature, electron phase space is available for the \ecapture \ daughter to undergo \bdecay, which occurs before any other nuclear reaction \citep{tsuruta1970}. As a result, the majority of Urca cooling pairs identified to date are odd-$A$ nuclei \citep{schatz2014}.

Here we focus on the complete identification of weaker Urca pairs in the $A<106$ mass range\textrm{---}the highest mass produced in the $rp$-process \citep{schatz2001} \textrm{---} that are located at shallow depths ($|Q_{\rm EC}| \lesssim 5\, \mathrm{MeV}$). These Urca cycles occur in the neutron star ocean, at column depths inferred for superburst ignition $y_{\rm ign} = P/g \approx (0.5 \textrm{--} 3) \times 10^{12} \, \mathrm{g \ cm^{-2}}$ \citep{cumming2006}. We identify a total of 85 odd-A isotopes that form Urca cycles in the neutron star's ocean. In addition to ground-state-to-ground-state transitions, we include $e^{-}$-captures onto excited states with excitation energies $E_{\rm X} \sim k_{\rm B}T\sim 100\, \mathrm{keV}$ for ocean temperatures near $T \sim 10^{9} \, \mathrm{K}$ reached in superbursting neutron stars \citep{cumming2004}.

For each odd-$A$ isotope, we calculate the total neutrino luminosity
from $e^{-}$-capture/$\beta^{-}$-decay cycles using analytic
expressions for the reaction rates \citep{tsuruta1970}. The
$ft$-values, which measure the transition strength of $e^{-}$-captures and $\beta^{-}$-decays, are based on experimental data\footnote{Evaluated Nuclear Structure Data File (ENSDF) \textrm{---} a computer file of experimental nuclear structure data maintained by the National Nuclear Data Center, Brookhaven National Laboratory ({\tt www.nndc.bnl.gov})\textrm{---}as of 2015 December 11} when available; otherwise, $ft$-values are obtained from
Table 1 of \citet{singh1998}, using the experimentally determined
ground state spin-parities $J^{\pi}$ of the $e^{-}$-capture parent and daughter nuclei. 

\begin{figure*}
\centering
\includegraphics{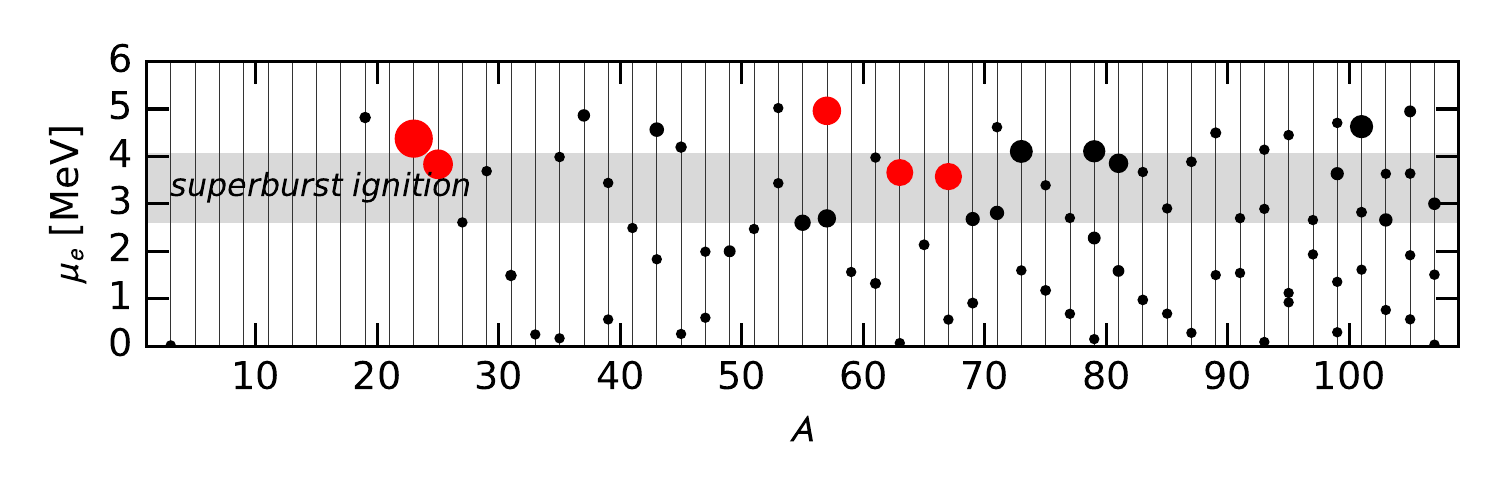}
\vspace{-0.4cm}
\caption{Depth of Urca pairs of a given mass number $A$. The size of data points corresponds to the neutrino luminosity of the Urca pair (the coefficient $L_{34}$ given in Equation~\ref{eq.l34}). Urca pairs with $X \cdot L_{34} \geq 5$ are colored in red for $X\,=\,1$. The gray band indicates empirical constraints on the superburst ignition depth between $y_{\rm ign} \approx 0.5 \textrm{--} 3 \times 10^{12} \, \mathrm{g \ cm^{-2}}$ \citep{cumming2006}.}
\label{figure.Lnu}
\end{figure*}

\begin{table}
\caption{Ocean Urca pairs \label{table.ocean_pairs}}
\centering
\renewcommand{\arraystretch}{1.4}
\begin{tabular}{lccr}
\hline
\hline

\textrm{Urca Pair, ${^A_Z}X$} & $|Q_{\rm EC}|$\,(MeV) & $y_{12} \footnotemark[1]$ & $X\cdot L_{34}\footnotemark[2]$ \\
\hline
$^{81}_{35}\mathrm{Br}\,\textrm{--}\, ^{81}_{34}\mathrm{Se}$ & $ 1.59$  & $0.1$ & $<0.1$  \\ 
$^{49}_{22}\mathrm{Ti}\,\textrm{--}\, ^{49}_{21}\mathrm{Sc}$ & $ 2.00$  & $0.2$ & $0.1$\\ 
$^{65}_{29}\mathrm{Cu}\,\textrm{--}\, ^{65}_{28}\mathrm{Ni}$ & $ 2.14$  & $0.2$ & $<0.1$ \\ 
$^{55}_{25}\mathrm{Mn}\,\textrm{--}\, ^{55}_{24}\mathrm{Cr}$ & $ 2.60$  & $0.4$ & $0.8$ \\ 
$^{69}_{30}\mathrm{Zn}\,\textrm{--}\, ^{69}_{29}\mathrm{Cu}$ & $ 2.68$  & $0.5$ & $0.3$ \\ 
$^{57}_{26}\mathrm{Fe^*}\,\textrm{--}\, ^{57}_{25}\mathrm{Mn}$ & $ 2.70$  & $0.5$ & $1.4$ \\ 
$^{67}_{29}\mathrm{Cu}\,\textrm{--}\, ^{67}_{28}\mathrm{Ni}$ & $ 3.58 $ & $1.5$ & $6.3$ \\ 
$^{63}_{28}\mathrm{Ni^*}\,\textrm{--}\, ^{63}_{27}\mathrm{Co}$ & $ 3.66 $ & $1.7$ & $6.1$ \\ 
$^{25}_{12}\mathrm{Mg}\,\textrm{--}\,^{25}_{11}\mathrm{Na}$ & $3.83 $ & $2.0$ & $8.2$ \\ 
$^{81}_{34}\mathrm{Se}\,\textrm{--}\,^{81}_{33}\mathrm{As}$ & $ 3.86$ & $2.1$ & $1.8$\\ 
$^{73}_{31}\mathrm{Ga}\,\textrm{--}\, ^{73}_{30}\mathrm{Zn}$ & $4.11$ & $2.7$ & $3.5$\\ 
$^{79}_{33}\mathrm{As}\,\textrm{--}\, ^{79}_{32}\mathrm{Ge}$ & $4.11$ & $2.7$ & $3.1$\\ 
$^{23}_{11}\mathrm{Na}\,\textrm{--}\,^{23}_{10}\mathrm{Ne}$ & $4.38 $ & $3.4$ & $17$\\ 
$^{101}_{42}\mathrm{Mo^*}\,\textrm{--}\,^{101}_{41}\mathrm{Nb}$ & $4.63$ & $4.3$ & $3.6$ \\ 
$^{57}_{25}\mathrm{Mn}\,\textrm{--}\,^{57}_{24}\mathrm{Cr}$ & $4.96$ & $5.7$ & $7.2$\\ 
\hline
\end{tabular}
\footnotetext[1]{$y_{12} \equiv y/(10^{12} \, \mathrm{g \ cm^{-2}})$, calculated with $g_{14} = 1.85$.}
\footnotetext[2]{Calculated with $X \, = \, 1$.}
 \end{table}

Following the approach of \citet{tsuruta1970}, in the relativistic limit ($\mu_e \approx E_{\rm F} \gg m_e c^2$) the specific neutrino emissivities from $e^{-}$-captures ($+$) and \bdecays \ ($-$) on nuclei in the degenerate electron ocean are
\begin{equation}
\epsilon_{\nu}^{\pm} \approx m_e c^2 \left(\frac{\mathrm{ln} \, 2}{ft} \right) {\langle F \rangle^{\pm} n^{\pm}} I^{\pm}(E_{\rm F}, T) \ ,
\end{equation}

\noindent where $n^{\pm}$ is the number density of nuclei, the Coulomb factor can be expressed as $\langle F \rangle^{\pm} \approx 2\pi \alpha Z / |1-\exp{(\mp 2\pi \alpha Z)}|$, and $\alpha \approx 1/137$ is the fine-structure constant. The electron phase space integrals are
\begin{equation}
I^{+}(E_{\rm F}, T) = \int^{\infty}_{W_m} {W\sqrt{W^2 - 1}(W-W_m)^3 S dW} \ ,
\end{equation}
\begin{equation}
I^{-}(E_{\rm F}, T) = \int^{W_m}_1  {W\sqrt{W^2 - 1} (W_m - W)^3  (1-S) dW}\ ,
\end{equation}

\noindent where $W \equiv E_e/m_e c^2$ is the electron energy, $W_m \equiv |Q_{\rm EC}|/m_e c^2$, $S = ({1 + \exp{[(E_e-E_{\rm F})/k_{\rm B}T}]})^{-1}$ is the statistical factor, $E_{\rm F}\approx 3.7\, {\rm MeV}\ (\rho_9Y_e/0.4)^{1/3}$ is the electron Fermi energy, $\rho_9 \equiv \rho/(10^9 \, \mathrm{g \ cm^{-3}}) $, and $Y_e \approx Z/A$ is the electron fraction. The total neutrino luminosity in the Urca reaction shell is found by integrating the sum of the specific neutrino emissivities from \ecaptures \ and \bdecays \ over the shell,
\begin{equation} \label{equation.lnu}
L_{\nu} \approx {4 \pi R^2} \int_{\rm shell} (\epsilon_{\nu}^+ + \epsilon_{\nu}^-)\, dz'  \ ,
\end{equation}

\noindent where $R$ is the radius of the neutron star. The Urca cycle occurs over a thin shell defined by the temperature $(\Delta R)_{\rm shell} \approx Y_e{k_{\rm B}T}/{m_u g}$ \citep{cooper2009,schatz2014}, where $(\Delta R)_{\rm shell} \ll R$, and the surface gravity of the neutron star is $g = GM/R^2$. The integral in Equation~\ref{equation.lnu} can be solved analytically by transforming the integration variable $dz' \approx (dP/d\mu_e)(d\mu_e/\rho g)$, where the pressure from degenerate electrons is $P\approx 3.6 \times 10^{26} \, \mathrm{ergs \ cm^{-3}} (\rho_9 Y_e/0.4)^{4/3}$. The neutrino luminosity from the Urca shell can be expressed as
\begin{equation}
L_{\nu} \approx L_{34} \times 10^{34} \, \mathrm{ergs \ s^{-1}} \,  X T_9^{5} \left( \frac{g_{14}}{2} \right)^{-1} R_{10}^2 \ ,
\end{equation}

\noindent where the parameter $L_{34}$ is a function of nuclear properties only,
\begin{equation}  \label{eq.l34}
L_{34} =  0.87 \,  \left( \frac{10^6 \, \mathrm{s}}{ft} \right) \left(\frac{56}{A} \right) \left( \frac{Q_{\rm EC}}{4 \, \mathrm{MeV}} \right)^{5}  \left( \frac{\langle F \rangle^*}{0.5}\right) \ ,
\end{equation}
\noindent and we define the parameters: $T_9 \equiv T/(10^9
\, \mathrm{K})$ is the ocean temperature, $g_{14} \equiv g/(10^{14} \, \mathrm{cm \
s^{-2}})$, $R_{10} \equiv R/(10 \, \mathrm{km})$, $ \langle F \rangle^* \equiv \langle F \rangle^+ \langle F \rangle^- /(\langle F \rangle^+  + \langle F \rangle^-)$, and $X \equiv A m_u n/\rho$ is the
mass fraction of the parent nucleus in the composition. 

Urca pairs are formed by all odd-$A$ nuclei studied, as can be seen in Figure~\ref{figure.Lnu}, where Urca cooling pairs are shown in relation to observed superburst ignition depths in a neutron star with $g_{14} = 1.85$. The 15 strongest Urca pairs in the superburst ignition region are shown in
Table~\ref{table.ocean_pairs}. The strongest cooling pairs occur at depths $y \gtrsim 10^{11} \, \mathrm{g \ cm^{-2}}$, and five pairs have $X\cdot L_{34} \geq 5$ for $X\,=\,1$. The strongest cooling pair is $^{23}$Na\textrm{--}$^{23}$Ne, which Urca-cycles near $y \approx 3.4 \times 10^{12} \, \mathrm{g \ cm^{-2}}$ and has $X \cdot L_{34}=17$. The strongest Urca pair formed from an excited state is $^{63}_{28}\mathrm{Ni^*}\,\textrm{--}\, ^{63}_{27}\mathrm{Co}$, which has $X \cdot L_{34}=6.1$ and Urca-cycles near $y \approx 1.7 \times 10^{12} \, \mathrm{g \ cm^{-2}}$.

\section{Urca cooling and carbon ignition} \label{s.superburst}

Urca pair neutrino cooling may impact the ignition conditions for thermonuclear bursts. In particular, Urca cooling lowers the ocean's steady-state temperature and makes superbursts ignite deeper. In this section, we explore the effect of Urca cooling's depth and strength on unstable \cc \ ignition. 

We calculate unstable carbon ignition in an ocean with an iron mass fraction $X_{\rm Fe} = 0.8$ and a carbon mass fraction $X_{\rm C} = 0.2$ following the approach of \citet{potekhin2012} using a modified one-zone approximation that reproduces the results of the stationary point method. \citet{potekhin2012} argued that this is a more appropriate way to calculate the ignition criterion than the local criterion used by previous authors \citep{cumming2001,cumming2006}. The local cooling rate from thermal diffusion is approximated by $\epsilon_{\rm cool} = \rho K_{\rm eff} T/ y^2$ \citep{fujimoto1981}, where $K_{\rm eff} \approx 0.16\,K$ is the effective thermal conductivity in the modified one-zone approximation. For the local heating rate from carbon burning, we use the \cc \ reaction rate given by \citet{yakovlev2010} that includes electron screening and corrections due to a multicomponent plasma \citep{potekhin2000,potekhin2009}, as well as quantum corrections to the rate \citep{potekhin2012}. 

The quantum corrections to the \cc \ reaction rate become important when the classical closest approach distance for the two nuclei becomes larger than the inter-ionic spacing \citep{alastuey1978}. This can be measured by the ratio $3 \Gamma / \tau$, where $\Gamma = (Ze)^2/a k_{\rm B} T$ is the Coulomb coupling parameter, $a = (3/4\pi n_{\rm ion})^{1/3}$ is the inter-ionic spacing, and $\tau = 3 E_{\rm pk} / k_{\rm B} T$ is the Gamow peak energy in units of $k_{\rm B} T$ \citep{salpeter1969}. We find that quantum effects are important at column depths $y \gtrsim 1.7 \times 10^{12} \, \mathrm{g \ cm^{-2}}$.

\subsection{Ocean Urca pairs} \label{sec.ocean_pairs_ign}

We calculate the neutron star ocean's thermal evolution using the open source code \dStar \ \citep{dstar}, which solves the general relativistic heat diffusion equation using the \mesa\ numerical library \citep{paxton2011, paxton2013, paxton2015}. During active accretion, crustal heating deposits $\approx 1.5 \, \mathrm{MeV}$ per accreted nucleon in the inner crust, which we place over the range $y= 5\times 10^{15}$ to $2 \times 10^{17}\, \mathrm{g \ cm^{-2}}$. To reproduce observed ignition depths, we add extra heating proportional to the accretion rate $Q_{\rm shallow}$ to the outer crust between $y=1.7\times 10^{15}$ and $5.4 \times 10^{15} \, \mathrm{g \ cm^{-2}}$ and the flux entering the ocean $F_b$ is taken at $y=1.7 \times 10^{15} \, \mathrm{g \ cm^{-2}}$. The net heat flux from crustal heating and extra heating entering the ocean is $F_b=Q_b \dot{m}$ \citep{cumming2006}, and $Q_b \approx 0.25 \, \mathrm{MeV} (\dot{m}/0.3\,\dot{m}_{\rm Edd})^{-1} \, $ gives carbon ignition at $y=5 \times 10^{11}\,\mathrm{g \ cm^{-2}}$, where $\dot{m}_{\rm Edd} = 8.8 \times 10^{4} \, \mathrm{g \ cm^{-2} \ s^{-1}}$ is the local Eddington mass accretion rate. The total luminosity entering the ocean is then $L_b = 10^{35} \, (\dot{M}/10^{17} \, \mathrm{g\ s^{-1}}) (Q_b / 10^{18} \, \mathrm{erg \ g^{-1}})$, where $\dot{M}$ is the global mass accretion rate. Therefore, the Urca cooling luminosity begins to balance the heating luminosity entering the ocean when $X\cdot L_{34} \approx 130\, (0.6/T_9)^5  (\dot{M}/10^{17} \, \mathrm{g\ s^{-1}}) (Q_b / 10^{18} \, \mathrm{erg \ g^{-1}})$. 


\begin{figure}
\centering
\includegraphics[width=0.95\apjcolwidth]{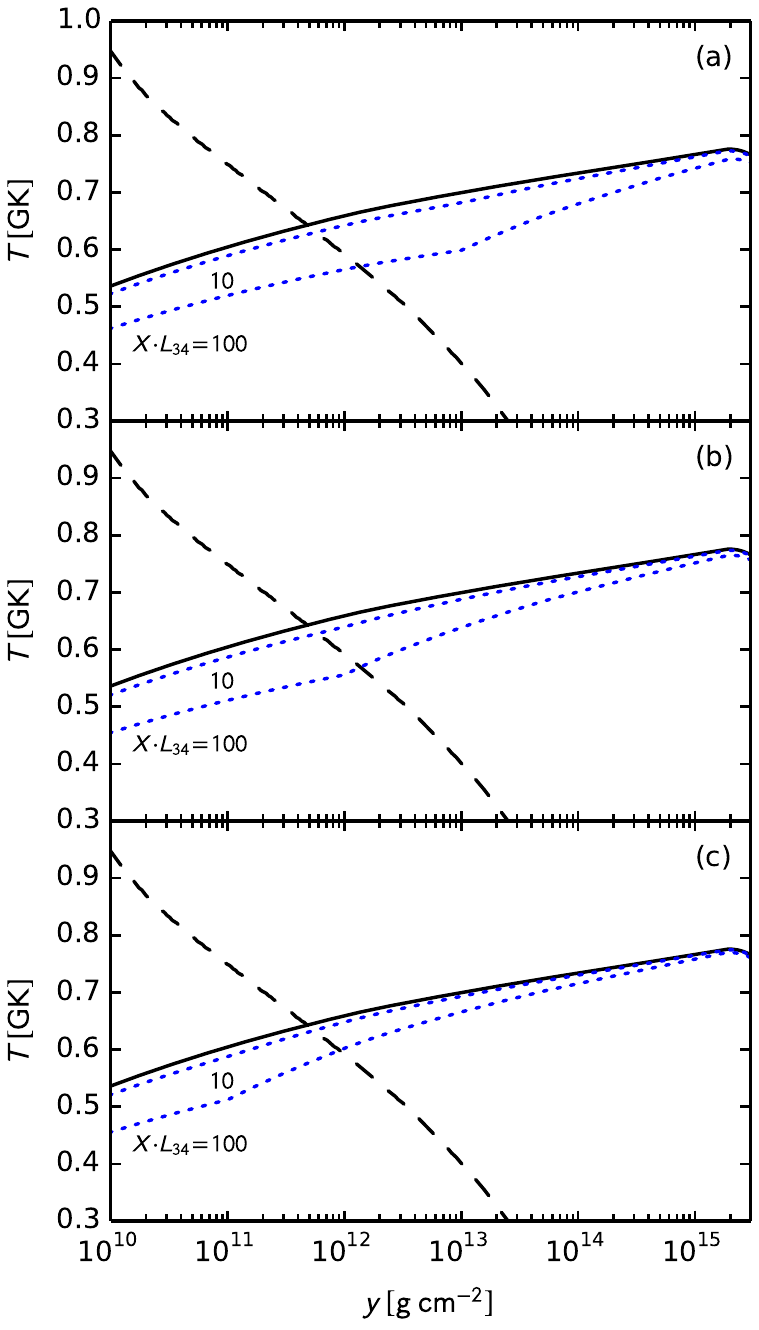}
\vspace{-0.1cm}
\caption{Steady-state ocean temperature as a function of column depth for a neutron star with $\dot{m} = 0.3 \, {\dot{m}_{\rm Edd}}$, $M = 1.4 \, \mathrm{M_{\odot}}$, $R=10\,\mathrm{km}$, and $T_{\rm core} = 3 \times 10^{7} \, \mathrm{K}$. The dashed black curves indicate the ignition of unstable carbon burning in a mixed iron-carbon ocean with $X_{\rm Fe} = 0.8$ and $X_{\rm C}= 0.2$. The solid black curves indicate ocean's without Urca cooling layers. In each panel, the lower dotted blue curve is for $X \cdot L_{34} =100$ and the upper dotted blue curve is for $X \cdot L_{34} = 10$, for Urca pairs located at (a)  $y=10^{13} \, \mathrm{g \ cm^{-2}}$, (b) $y= 10^{12} \, \mathrm{g \ cm^{-2}}$, and (c) $y=10^{11} \, \mathrm{g \ cm^{-2}}$. }
\label{fig:ignition}
\end{figure}

We investigate the thermal evolution in a neutron star with mass $M = 1.4 \, \mathrm{M_{\odot}}$, radius $R=10\,\mathrm{km}$, and core temperature $T_{\rm core} = 3 \times 10^{7} \, \mathrm{K}$. The neutron star accretes at a local accretion rate $\dot{m} = 0.3\, \dot{m}_{\rm Edd}$ \textrm{---} the minimum accretion rate to ignite carbon in an ocean with a carbon mass fraction $X_{\rm C} = 0.2$ \citep{cumming2006} \textrm{---} until the ocean temperature reaches steady state. The model requires $Q_{\rm shallow} \approx 5 \, \mathrm{MeV}$ per accreted nucleon to provide $Q_b = 0.25 \, \mathrm{MeV}$ per accreted nucleon into the ocean, which gives ignition at $y_{\rm ign,0}=5\times 10^{11} \, \mathrm{g \ cm^{-2}}$ for this accretion rate. A light-element column of hydrogen and helium extends from the neutron star surface to a depth $y=10^{8} \, \mathrm{g \ cm^{-2}}$ \citep{brown1998}. 

To examine the impact of Urca cooling's location on the ocean temperature, we run three models with different Urca pair locations at $y/y_{\rm ign, 0} = 0.2, 2, 20$. We test the two values $X \cdot L_{34} = 10, 100$ for each Urca pair to represent the characteristic strength of ocean Urca pairs (see Table~\ref{table.ocean_pairs}). Figure~\ref{fig:ignition} shows the ocean's steady-state temperature during accretion for the three different Urca pair locations. 

Neutrino emission from ocean Urca pairs lowers the ocean's steady-state temperature; for example, at the depth of the Urca pair the temperature decreases by $\approx 1 \%$ ($\approx 10 \%$) for $X \cdot L_{34} = 10$ ($X\cdot L_{34} = 100$). Furthermore, carbon ignition occurs deeper in the cooler ocean. For example, an Urca pair located at $y/y_{\rm ign, 0} = 2$ moves ignition to $y_{\rm ign}\approx 1.2 \, y_{\rm ign, 0}$ ($y_{\rm ign}\approx 2.8  \, y_{\rm ign,0}$) for $X \cdot L_{34} =10$ ($X \cdot L_{34} = 100$). The results are similar for the other Urca pair locations tested. When an Urca pair is located at $y/y_{\rm ign,0} = 0.2$ the carbon ignition depth increases to $y_{\rm ign}\approx 1.1 \, y_{\rm ign, 0}$ ($y_{\rm ign}\approx 1.8 \, y_{\rm ign,0}$) for an Urca pair with $X \cdot L_{34} =10$ ($X \cdot L_{34} = 100$). When an Urca pair is located at $y/y_{\rm ign,0} = 20$ the carbon ignition depth increases to $y_{\rm ign}\approx 1.3 \, y_{\rm ign, 0}$ ($y_{\rm ign}\approx 2.8  \, y_{\rm ign,0}$) for an Urca pair with $X \cdot L_{34} =10$ ($X \cdot L_{34} = 100$). 

Ocean Urca pairs reduce the ocean's steady-state temperature and require deeper carbon ignition only when $X \cdot L_{34} \gtrsim 10$. At the abundances expected for X-ray burst and superburst burning \citep{schatz2014}, however, all ocean Urca pairs have $X \cdot L_{34} \ll 1$. Furthermore, the total neutrino cooling from all ocean pairs $\sum_i {X_i \cdot L_{34, i}} \approx 0.1$ for X-ray burst ashes and $\approx 1.6$ for superburst ashes. Moreover, a majority of ocean pairs have experimentally measured $ft$-values, and the resulting uncertainty in $L_{34}$ is less than a factor of $\approx 2$ and insufficient to allow $X \cdot L_{34} > 1$ for any ocean Urca pair. Therefore, we expect an ocean with Urca pairs to have a steady-state temperature and carbon ignition depth consistent with an ocean without Urca pairs. 

\subsection{Crust Urca pairs}

The neutrino luminosity from an Urca pair scales as $|Q_{\rm EC}|^5$, which means that crust Urca pairs are a factor of $\sim 100 \textrm{--} 1000$ more luminous than ocean pairs. Furthermore, for the abundances of X-ray burst ashes and superburst ashes \citep{schatz2014}, crust Urca pairs have between $X\cdot L_{34} \sim 10\textrm{--}100$, as can be seen in Table~\ref{table.crust_pairs}. As a consequence, crust pairs are more likely to be important coolers at the temperatures of interest near $\sim 10^9 \,\mathrm{K}$ because their neutrino luminosities are comparable to the heating luminosity entering the ocean $L_b = 10^{35} \, (\dot{M}/10^{17} \, \mathrm{g\ s^{-1}}) (Q_b / 10^{18} \, \mathrm{erg \ g^{-1}})$, which requires an Urca pair neutrino luminosity near $X\cdot L_{34} \approx 130\, (0.6/T_9)^5  (\dot{M}/10^{17} \, \mathrm{g\ s^{-1}}) (Q_b / 10^{18} \, \mathrm{erg \ g^{-1}})$.  Although crust Urca pairs are located at depths $y \gg y_{\rm ign}$, they lower the ocean's steady-state temperature by limiting the net heat flux entering the ocean from crustal and extra heating. 

\begin{table}
\caption{Crust Urca pairs \label{table.crust_pairs}}
\centering
\renewcommand{\arraystretch}{1.4}
\begin{tabular}{lccrr}
\hline
\hline

\textrm{Urca Pair, ${^A_Z}X$} & $|Q_{\rm EC}|$\,(MeV) & $y_{14} \footnotemark[1]$ & $X_{\rm XRB}\cdot L_{34}$ & $X_{\rm SB}\cdot L_{34}$ \\
\hline
$^{59}_{25}\mathrm{Mn}\,\textrm{--}\, ^{59}_{24}\mathrm{Cr}$ & $ 7.6$  & $0.3$ & $<0.1$ & $<0.1$ \\ 
$^{57}_{24}\mathrm{Cr}\,\textrm{--}\, ^{57}_{23}\mathrm{V}$ & $ 8.3$  & $0.4$ & $<0.1$ & $<0.1$ \\ 
$^{65}_{26}\mathrm{Fe}\,\textrm{--}\, ^{65}_{25}\mathrm{Mn}$ & $ 10.3$  & $1.1$ & $<0.1$ & $<0.1$ \\ 
$^{57}_{23}\mathrm{V}\,\textrm{--}\, ^{57}_{22}\mathrm{Ti}$ & $ 10.7$  & $1.2$ & $0.4$ & $10$ \\ 
$^{65}_{25}\mathrm{Mn}\,\textrm{--}\, ^{65}_{24}\mathrm{Cr}$ & $ 11.7$  & $1.8$ & $<0.1$ & $<0.1$ \\ 
$^{31}_{13}\mathrm{Al}\,\textrm{--}\, ^{31}_{12}\mathrm{Mg}$ & $ 11.8$  & $1.8$ & $1.2$ & $<0.1$ \\ 
$^{55}_{21}\mathrm{Sc}\,\textrm{--}\, ^{55}_{20}\mathrm{Ca}$ & $ 12.1$  & $2.0$ & $0.9$ & $200$  \\ 
$^{29}_{12}\mathrm{Mg}\,\textrm{--}\, ^{29}_{11}\mathrm{Na}$ & $ 13.3$  & $2.9$ & $1.0$ & $0.3$  \\ 
$^{33}_{13}\mathrm{Al}\,\textrm{--}\, ^{33}_{12}\mathrm{Mg}$ & $ 13.4$  & $3.0$ & $19$ & $0.5$ \\ 
$^{63}_{24}\mathrm{Cr}\,\textrm{--}\, ^{63}_{23}\mathrm{V}$ & $ 14.7$  & $4.4$ & $<0.1$ & $<0.1$ \\ 
\hline
\end{tabular}
\footnotetext[1]{$y_{14} \equiv y/(10^{14} \, \mathrm{g \ cm^{-2}})$, calculated with $g_{14} = 1.85$.}
\end{table}

To determine the effect of crust Urca pairs on the ocean temperature, we first calculate $X\cdot L_{34}$ values for the strongest Urca pairs identified in the crust \citep{schatz2014}. The mass fractions of Urca pairs are taken from multizone nuclear reaction network calculations of X-ray burst and superburst nuclear burning \citep{schatz2014}. The crust Urca pairs, along with their location and $X\cdot L_{34}$ values, are shown in Table~\ref{table.crust_pairs}. We place all Urca pairs with $X\cdot L_{34} > 0.1$ in the same neutron star thermal evolution model used in Section~\ref{sec.ocean_pairs_ign}.

\begin{figure}
\centering
\includegraphics[width=0.95\apjcolwidth]{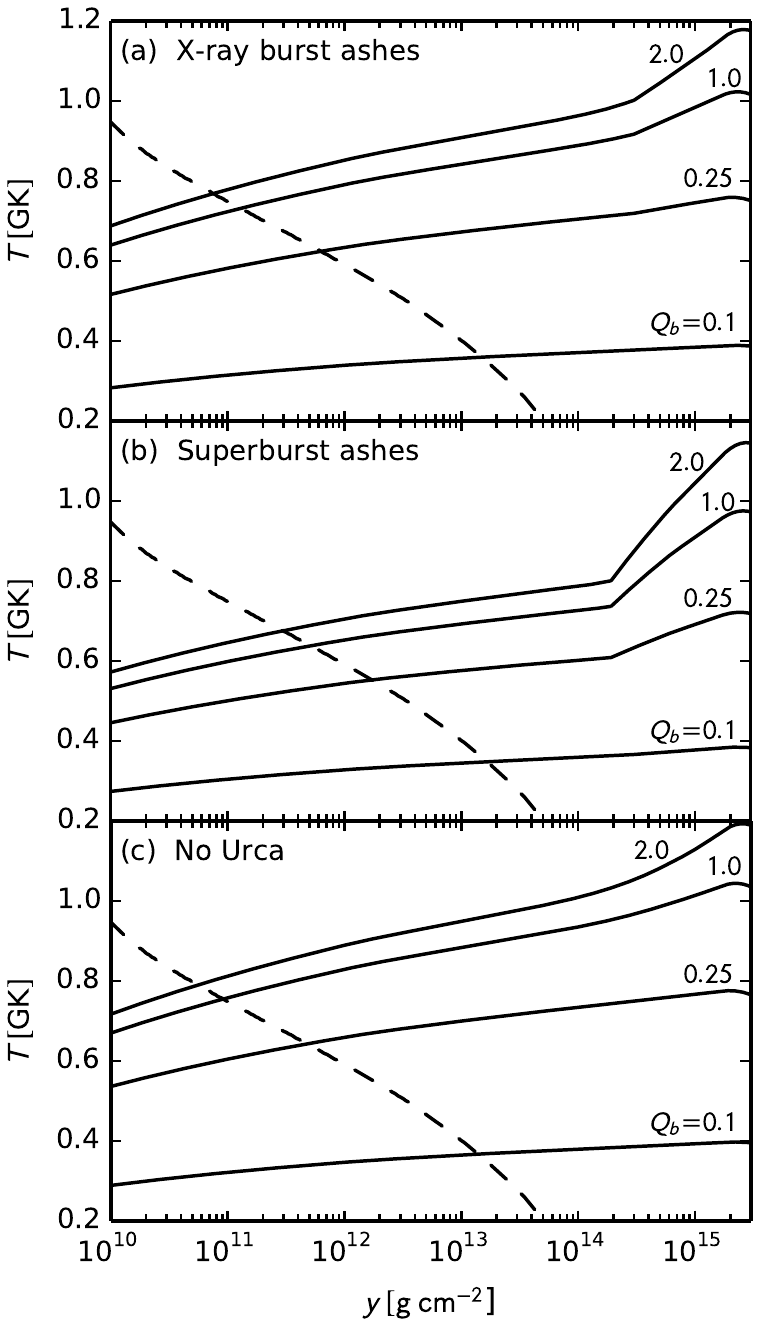}
\vspace{-0.2cm}
\caption{Steady-state ocean temperature as a function of column depth for a neutron star model with $\dot{m} = 0.3 \, {\dot{m}_{\rm Edd}}$, $M = 1.4 \, \mathrm{M_{\odot}}$, $R=10\,\mathrm{km}$, and core temperature $T_{\rm core} = 3 \times 10^{7} \, \mathrm{K}$. The dashed black curves indicate the ignition of unstable carbon burning in a mixed iron-carbon ocean with $X_{\rm Fe} = 0.8$ and $X_{\rm C}= 0.2$. The solid black curves indicate ocean temperature profiles for $Q_b = 0.1, 0.25, 1.0, 2.0 \, {\rm MeV}$. (a) Crust Urca pairs at the abundances expected in X-ray burst ashes. (b) Crust Urca pairs at the abundances expected in superburst ashes. (c) Ocean temperature profiles without Urca cooling in the crust.}
\label{fig.crust_urca}
\end{figure}

In Figure~\ref{fig.crust_urca} we show ocean steady-state temperature profiles for $Q_b=0.1,0.25,1.0,2.0\, \mathrm{MeV}$ per accreted nucleon including all crust pairs with $X\cdot L_{34} > 0.1$. Because $\approx 10\, \%$ of crustal heating enters the ocean during steady state \citep{brown2000}, these values of $Q_b$ represent respectively $Q_{\rm shallow} \approx 1, 5, 10, 15 \, \mathrm{MeV}$ per accreted nucleon of extra heating in the crust. These values of $Q_{\rm shallow}$ are representative of the observational constraints on shallow heating from neutron star transients; for example, $\approx 1\, \mathrm{MeV}$ per accreted nucleon is needed in KS~1731--260 and MXB~1659--29 \citep{brown09}, and $\approx 6\textrm{--}16\, \mathrm{MeV}$ per accreted nucleon is needed in MAXI~J0556--332 \citep{deibel2015}. 

As can be seen in Figure~\ref{fig.crust_urca}, crust Urca pairs lower the ocean's steady-state temperature by removing $\approx 80-90 \%$ of the crustal heat flux $F_b$ entering the ocean. The ocean temperature is limited to $T \lesssim  10^9 \, \mathrm{K}$ ($T \lesssim 8 \times 10^8 \, \mathrm{K}$) at $y \lesssim 10^{14} \, \mathrm{g \ cm^{-2}}$ for X-ray burst ashes (superburst ashes). The carbon ignition depth increases when crust pairs are present, for all values of $Q_b$. For example, for a heat flux $Q_b = 0.25\, \mathrm{MeV}$ per accreted nucleon carbon ignition occurs at $y_{\rm ign} = 5 \times 10^{11} \, \mathrm{g \ cm^{-2}}$ without Urca cooling, whereas carbon ignition occurs near $y_{\rm ign} \approx 6.4 \times 10^{11} \, \mathrm{g \ cm^{-2}}$ ($y_{\rm ign} \approx 1.8 \times 10^{12} \, \mathrm{g \ cm^{-2}}$) when crust Urca pairs from X-ray burst ashes (superburst ashes) are present. The largest crustal heat flux $Q_b = 2 \, \mathrm{MeV}$ per accreted nucleon gives the shallowest ignition depth near $y_{\rm ign} \approx 7.8 \times 10^{10} \, \mathrm{g \ cm^{-2}}$ ($y_{\rm ign} \approx 3.0 \times 10^{11} \, \mathrm{g \ cm^{-2}}$) for X-ray burst pairs (superburst pairs). 

\begin{figure*}
\centering
\includegraphics[width=0.95\apjcolwidth]{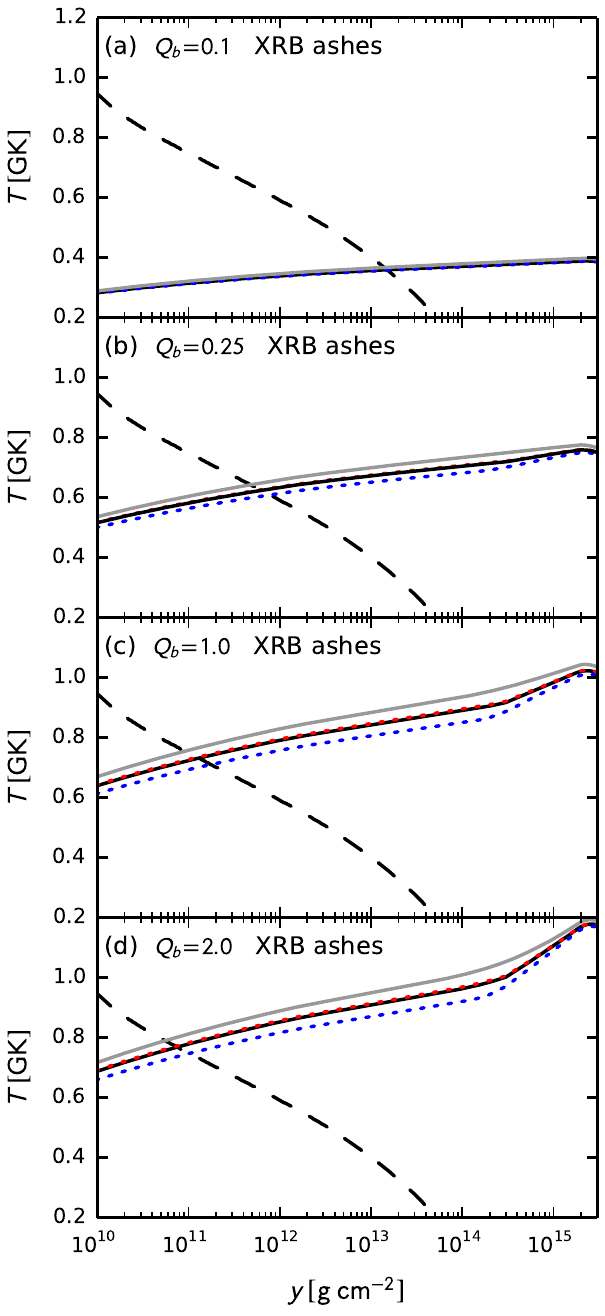}
\includegraphics[width=0.95\apjcolwidth]{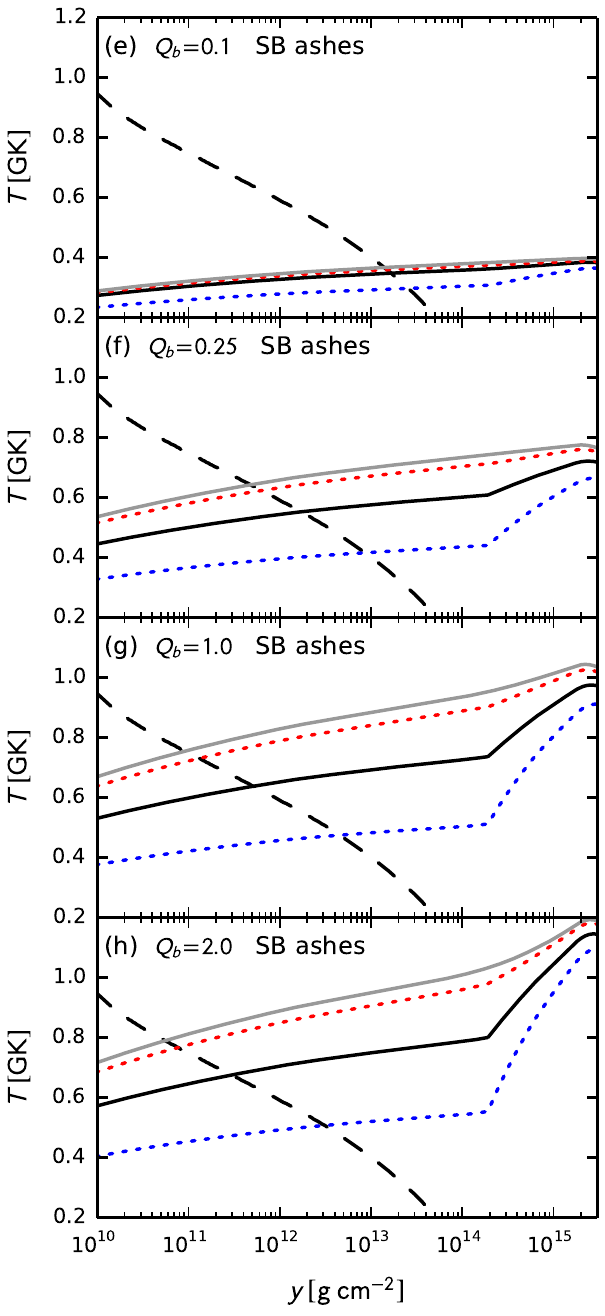}
\vspace{-0.1cm}
\caption{Steady-state ocean temperature as a function of column depth for different values of $Q_b = 0.1, 0.25, 1.0, 2.0 \, {\rm MeV}$ (black solid curves). Panels (a)\textrm{--}(d) have $X \cdot L_{34}$ values calculated for the abundances of X-ray burst ashes (XRB), and panels (e)\textrm{--}(h) have $X \cdot L_{34}$ values calculated for the abundances of superburst ashes (SB). The upper dashed red curves are a factor of 10 enhancement of the $ft$-value, and the lower dashed blue curves are a factor of 10 reduction in the $ft$-value. The solid gray curves are the temperature profiles without Urca cooling pairs. The dashed black curve indicates the ignition of unstable carbon burning in a mixed iron-carbon ocean with $X_{\rm Fe} = 0.8$ and $X_{\rm C}= 0.2$.}
\label{fig.ignition_uncertainty}
\end{figure*}

\ \\
\subsection{The superburst in 4U~1636--536} 

The shallow ignition of the superburst in 4U~1636-536 (Wijnands 2001; Strohmayer \& Markwardt 2002; Keek et al. 2014a,b) near $y_{\rm ign} \approx 2 \times 10^{11} \, \mathrm{g \ cm^{-2}}$ \citep{keek2015} is consistent with a crust made of X-ray burst ashes and $Q_{b} \approx 2 \, \mathrm{MeV}$ per accreted nucleon, as can be seen in Figure~\ref{fig.crust_urca}. Within the uncertainties, enhanced \ecapture/\bdecay $ft$-values allow carbon ignition depths consistent with the superburst ignition depth in 4U~1636--536 for smaller values of $Q_b$ than required in the mid-range $ft$-value case in a crust composed of superburst ashes. 


Crust Urca pairs are composed of more neutron-rich nuclei than ocean pairs. These neutron-rich nuclei lie far from stability, and few have measured $ft$-values. As a consequence, crust Urca pairs have at least a factor of $\approx 10$ uncertainties in the transition strength $ft$-value and thereby a factor of $\approx 10$ uncertainty in their neutrino luminosities. Note that similar uncertainties near a factor of $\approx 10$ are possible in the calculation of $X$ due to reaction rate uncertainties \citep{parikh2008}. For the crust pairs without measured $ft$-values, we now examine the neutrino luminosities within the uncertainty in the $ft$-value. In Figure~\ref{fig.ignition_uncertainty} we show ocean temperature profiles for $X \cdot L_{34}$ for the abundances in X-ray burst and superburst ashes, for a factor of 10 reduction and a factor of 10 enhancement in the $ft$-value. Note that a reduction in the $ft$-value increases the neutrino luminosity of the Urca pair (see Equation~\ref{eq.l34}).

A crust composed of X-ray burst ashes and a value of $Q_b \approx 1\, \mathrm{MeV}$ per accreted nucleon allow carbon ignition near $y_{\rm ign} \approx 2 \times 10^{11} \, \mathrm{g \ cm^{-2}}$ for $ft$-values calculated using our method. Taking into account the uncertainty in $ft$-values, carbon ignition at this depth is allowed for $Q_b \gtrsim 0.75 \, \mathrm{MeV}$ per accreted nucleon if $ft$-values are enhanced by a factor of 10 in X-ray burst ashes, as can be seen in Figure~\ref{fig.ignition_uncertainty}. 

In a crust composed of superburst ashes, even a large value of extra heating $Q_{b} = 2 \, \mathrm{MeV}$ per accreted nucleon is not consistent with carbon ignition below $y_{\rm ign} \lesssim 3.0 \times 10^{11} \, \mathrm{g \ cm^{-2}}$. When uncertainties in $ft$-values are taken into account, however, carbon ignition at $y_{\rm ign} = 2 \times 10^{11} \, \mathrm{g \ cm^{-2}}$ is allowed for $Q_b \gtrsim 1 \, \mathrm{MeV}$ per accreted nucleon if $ft$-values are enhanced by a factor of 10 in superburst ashes.

In addition to nuclear physics uncertainties, a shallower extra heat source can reconcile carbon ignition with the inferred superburst ignition depth in 4U~1636--536. In the previous sections, we place the extra heating at depths greater than crust Urca pairs ($y \gtrsim 2 \times 10^{15} \, \mathrm{g \ cm^{-2}}$) following the approach of previous carbon ignition models \citep{brown1998,cumming2001}. The extra heating may be shallower, however, as is the case in neutron star transients. For example, extra heating in KS~1731--260 and MXB~1659--29 must be  $y \lesssim 2 \times 10^{14} \, \mathrm{g \ cm^{-2}}$ to fit quiescent light curves, and in MAXI~J0556--332 the inferred extra heating is in the range $y\approx 2 \times 10^{13} \textrm{--} 2 \times 10^{14} \, \mathrm{g \ cm^{-2}}$ \citep{deibel2015}.

\begin{figure}
\centering
\includegraphics[width=0.95\apjcolwidth]{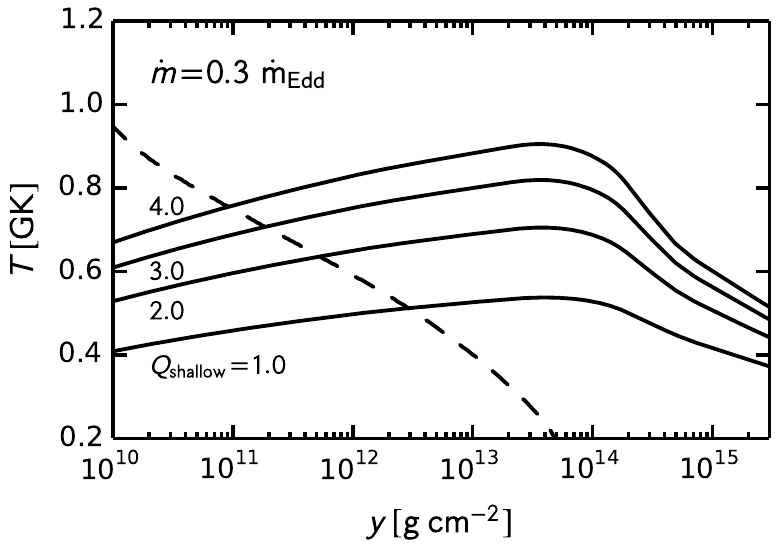}
\vspace{-0.2cm}
\caption{Steady-state ocean temperature as a function of column depth for a neutron star model with $\dot{m} = 0.3 \, {\dot{m}_{\rm Edd}}$, $M = 1.4 \, \mathrm{M_{\odot}}$, $R=10\,\mathrm{km}$, and core temperature $T_{\rm core} = 3 \times 10^{7} \, \mathrm{K}$. Solid black curves are steady-state temperature profiles for the heating strengths $Q_{\rm shallow} = 1, 2, 3,$ and $4\, \mathrm{MeV}$. The dashed black curve indicates the ignition of unstable carbon burning in a mixed iron-carbon ocean with $X_{\rm Fe} = 0.8$ and $X_{\rm C}= 0.2$. }
\label{fig.shallow_heating}
\end{figure}

To examine carbon ignition conditions with a shallower extra heat source, we place extra heating at the depth constrained in MAXI~J0556--322, in the range $y\approx 2 \times 10^{13} \textrm{--} 2 \times 10^{14} \, \mathrm{g \ cm^{-2}}$. Extra heating deposited in this region means that the ocean temperature at $y \lesssim 8 \times 10^{13} \, \mathrm{g \ cm^{-2}}$ is unaffected by neutrino cooling from crust Urca pairs and crust pairs do not impact the carbon ignition depth. Figure~\ref{fig.shallow_heating} shows ocean temperature profiles for $Q_{\rm shallow} = 1, 2, 3,$ and $4 \, \mathrm{MeV}$ per accreted nucleon. When extra heat is deposited at these depths, a heating strength of $Q_{\rm shallow} \approx 3 \, \mathrm{MeV}$ per accreted nucleon is needed for ignition at $y \approx 2 \times 10^{11} \, \mathrm{g \ cm^{-2}}$. 

\begin{figure}
\centering
\includegraphics[width=0.95\apjcolwidth]{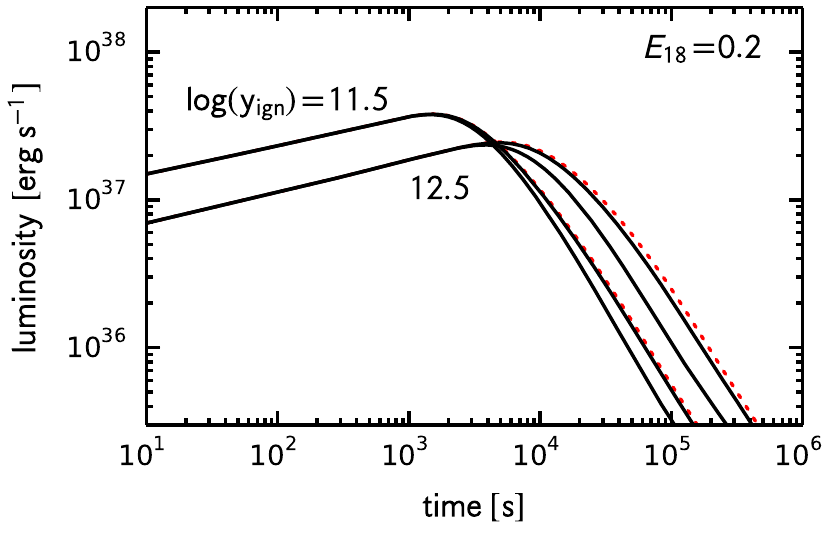}
\vspace{-0.2cm}
\caption{Cooling light curves for superbursts with $E_{18} = 0.2$, $\alpha=0.25$, $g_{14} = 1.53$, and ignition depths $y_{\rm ign}=3.2\times10^{11}\, \mathrm{g \ cm^{-2}}$ and $y_{\rm ign}=3.2\times10^{12}\, \mathrm{g \ cm^{-2}}$. The red dotted curves are models without Urca cooling. The solid curves are for a cooling source at $y=0.4\, y_{\rm ign}$ with $X \cdot L_{34} = 100,1000 $. }
\label{figure.superburst}
\end{figure}

\section{Urca cooling and superburst light curves} \label{s.lightcurve}

The strong $T^5$ dependence of the Urca pair neutrino luminosity means that ocean Urca pairs become more luminous after superburst ignition, when the ocean temperature reaches $T \gtrsim 10^{9} \, \mathrm{K}$ during carbon burning. To examine the impact of ocean Urca pairs after superburst ignition, we predict superburst light curves following the model of \citet{cumming2004}, which assumes that nuclear burning during the superburst occurs quickly and sets the initial temperature profile for the cooling light curve. The average energy input during the superburst is $E_{18} \equiv E_{\rm nuc}/(10^{18} \, \mathrm{ergs \ g^{-1}})$ and the initial cooling temperature profile is a power law with depth $T \propto y^{\alpha}$, where $y  \approx \, 1.8 \times 10^{12} \, \mathrm{g \ cm^{-2}}\rho_9^{4/3} (g_{14}/2)^{-1} (Y_e/0.4)^{4/3} $ is the column depth. We follow the thermal evolution from the initial temperature profile using the code \code{burstcool}\footnote{\code{https://github.com/andrewcumming/burstcool}}, which integrates the heat diffusion equation assuming constant gravity over the bursting layer, and the emergent flux is redshifted to an observer at infinity. 

To explore Urca cooling's effect on light curve predictions, we examine two superburst ignition depths, $y_{\rm ign} = 3.2 \times 10^{11} \, \mathrm{g \ cm^{-2}}$ and  $y_{\rm ign} = 3.2 \times 10^{12} \, \mathrm{g \ cm^{-2}}$, for a neutron star mass $M = 1.4 \, \mathrm{M_{\odot}}$, neutron star radius $R=11\,\mathrm{km}$, and $g_{14}=1.53$. An Urca cooling pair is placed in close proximity to the superburst ignition layer at $y= 0.4 \, y_{\rm ign}$ in both superbursts. Note that beneath the ignition layer ($y> y_{\rm ign} $) the temperature is $T \lesssim 5 \times 10^8 \, \mathrm{K}$ and cooling from ocean Urca pairs is small. Therefore, we only show models with an Urca pair located shallower than ignition ($y < y_{\rm ign}$) where the temperature is $T\gtrsim 10^9 \, \mathrm{K}$ at the outset of cooling. The cooling light curve predictions for these superbursts are shown in Figure~\ref{figure.superburst} for $X\cdot L_{34} =100$ and 1000. 

In the absence of Urca cooling pairs, the superburst with ignition depth $y_{\rm ign} = 3.2 \times 10^{11} \, \mathrm{g \ cm^{-2}}$ ($y_{\rm ign} = 3.2 \times 10^{12} \, \mathrm{g \ cm^{-2}}$) has a peak luminosity $L_{\rm peak} \approx 4 \times 10^{37} \, \mathrm{ergs \ s^{-1}}$ ($L_{\rm peak} \approx 2 \times 10^{37} \, \mathrm{ergs \ s^{-1}}$). Urca cooling only minimally affects the peak superburst luminosity because $L_{\nu} \ll L_{\rm peak}$, even for Urca pairs with $X \cdot L_{34} = 1000$, as can be seen in Figure~\ref{figure.superburst}. Urca cooling generally causes an earlier and more rapid decline of the superburst cooling light curve, similar to a light curve with a reduced ignition depth. At the abundances predicted for X-ray burst and superburst ashes, ocean pairs located at $y<y_{\rm ign}$ have $X \cdot L_{34} \ll 1$ and do not impact superburst cooling light curves. Moreover, ocean pairs with unmeasured $ft$-values do not have $X \cdot L_{34} > 1$ even with a factor of 10 reduction in the $ft$-value.

\section{Discussion}\label{s.discussion}

We have explored the neutrino luminosities of 85 Urca pairs in the neutron star ocean near the depths inferred for superburst ignition in the range $y_{\rm ign} \approx (0.5 \textrm{--} 3) \times 10^{12} \, \mathrm{g \ cm^{-2}}$ \citep{cumming2006}. We find that Urca cooling occurs throughout an  ocean composed of X-ray burst and/or superburst ashes. When combined with the crust Urca pairs previously identified \citep{schatz2014}, our results suggest that Urca cooling occurs at varying strengths throughout the entire outer crust at mass densities $10^{6} \lesssim \rho \lesssim 4 \times 10^{11} \, \mathrm{g \ cm^{-3}}$. 

At large enough abundances ocean Urca pairs have the potential to cool the ocean and increase carbon ignition depths. At the abundances calculated from X-ray burst and superburst nuclear burning \citep{schatz2014}, however, ocean pairs have weak neutrino luminosities ($X\cdot L_{34} \ll 1$) and do not impact carbon ignition depths. Furthermore, the neutrino luminosities of ocean Urca pairs located at $y \lesssim y_{\rm ign}$ are too weak to cause an early decline of the superburst light curve. That is, ocean pairs have no effect on the cooling light curve because their neutrino luminosities are small relative to the superburst peak luminosity ($L_{\nu} \ll L_{\rm peak}$) even in the hot post-superburst ocean near $T \gtrsim 10^9 \, \mathrm{K}$.

For the abundances predicted from X-ray burst and superburst nuclear burning, crust Urca pairs have strong neutrino luminosities ($\sum_i X_i \cdot L_{34, i} \approx 230$ for superburst ashes and $\sum_i X_i \cdot L_{34, i} \approx 25$ for X-ray burst ashes) and increase carbon ignition depths even though they cool at greater depths $y \gg y_{\rm ign}$. Crust Urca pairs lower the ocean's steady-state temperature by preventing a majority ($\approx 80\textrm{--}90\%$) of the net crustal and extra heating from entering the ocean, as shown in Figure~\ref{fig.crust_urca}. In a crust composed of X-ray burst ashes, the ocean temperature is limited to $T \lesssim 10^{9} \, \mathrm{K}$ in the carbon ignition region and ignition depths are limited to $y_{\rm ign} \gtrsim 7.8 \times 10^{10} \, \mathrm{g \ cm^{-2}}$. In a crust composed of superburst ashes, the ocean temperature is limited to $T \lesssim 8 \times 10^{8} \, \mathrm{K}$  and carbon ignition depths are limited to $y_{\rm ign} \gtrsim 3.0 \times 10^{11} \, \mathrm{g \ cm^{-2}}$. Once nuclear uncertainties are taken into account, allowed carbon ignition depths are $y_{\rm ign} \gtrsim 7.4 \times 10^{10} \, \mathrm{g \ cm^{-2}}$ ($y_{\rm ign} \gtrsim 7.8 \times 10^{10} \, \mathrm{g \ cm^{-2}}$) in a crust composed of X-ray burst (superburst) ashes.

Our results suggest that the cooling thermal component of the 2001 February superburst in 4U~1636--536 (Wijnands 2001; Strohmayer \& Markwardt 2002; Keek et al. 2014a,b) is likely not caused by Urca cooling in the ocean. The cooling light curve declines much faster after the superburst peak than model predictions (with uncertainties about the spectral model used to fit the data; see discussion in Keek et al. 2015). Because the light curve declines after the superburst peak, Urca cooling would need to be near the carbon ignition depth to explain the cooling trend \citep{keek2015}. For the ignition depth near that of the superburst in 4U~1636--536  ($y_{\rm ign} \approx 2 \times 10^{11} \, \mathrm{g \ cm^{-2}}$), only Urca cooling pairs present at depths $y \lesssim y_{\rm ign}$ will impact cooling predictions; however, these ocean pairs are typically weak with $X \cdot L_{34} \ll 1$ and do not effect cooling light curve predictions. Indeed, Urca cooling in the ocean only marginally effects light curve predictions even for $X \cdot L_{34} = 1000$, as can be seen in Figure~\ref{figure.superburst}. 

The shallow ignition depth in 4U~1636--536 near $y_{\rm ign} \approx 2 \times 10^{11} \, \mathrm{g \ cm^{-2}}$ \citep{keek2015} provides constraints on the strength of extra heating and the composition of the crust in this source. If extra heating is located at $y \gtrsim 10^{15} \, \mathrm{g \ cm^{-2}}$ \textrm{---} as is done in superburst ignition models\textrm{---}a minimum heating strength of $Q_{\rm shallow} \approx 7.5 \, \mathrm{MeV}$ per accreted nucleon is required to ignite carbon near $y \approx 2 \times 10^{11} \, \mathrm{g \ cm^{-2}}$ in a crust composed of X-ray burst ashes. If the crust contains superburst ashes, the minimum heating strength for ignition at this depth is $Q_{\rm shallow} \approx 10 \, \mathrm{MeV}$ per accreted nucleon, and \ecapture /\bdecay \ $ft$-values must be larger than those obtained using our method (see Section~\ref{s.pairs}). This may indicate that the crust in 4U~1636--536 is composed primarily of X-ray burst ashes because this case requires less extra heating and is consistent with Urca cooling neutrino luminosities using the best-fit $ft$-values. Furthermore, superburst ignition models find that several Type I X-ray bursts occur during the intervals between the more energetic superbursts \citep{keek2012}. X-ray burst ashes accumulate in the crust over the course of $\sim 100\textrm{--}1000 \, \mathrm{yr}$ in the more frequent Type I X-ray bursts, and cooling from Urca pairs in these ashes may prevail over cooling from pairs in the less abundant superburst ashes.

By contrast, extra heating may be deposited at shallower depths near $y \lesssim 2 \times 10^{14} \, \mathrm{g \ cm^{-2}}$ as is the case in neutron star transients with extra shallow heating, for example, KS~1731--260 and MXB~1659--29 \citep{brown09}, and MAXI~J0556--332 \citep{deibel2015}. When deposited at shallower depths, an extra heating strength of $Q_{\rm shallow} \approx 3 \, \mathrm{MeV}$ per accreted nucleon is required to have carbon ignition at depths consistent with the superburst ignition depth in 4U~1636--536, as can be seen in Figure~\ref{fig.shallow_heating}.  Note that a weaker shallow heat source is consistent with neutron star transients that have similar accretion rates in the range $\dot{m} \sim (0.1 \textrm{--} 0.3) \, {\dot{m}_{\rm Edd}}$, for instance, MXB~1659--29 requires $\approx 1 \, \mathrm{MeV}$ per accreted nucleon \citep{brown09}. When extra heating is located at $y \lesssim 2 \times 10^{14} \, \mathrm{g \ cm^{-2}}$ \textrm{---} shallower than most crust Urca pairs \textrm{---} no constraints on crust composition are possible because crust Urca pairs do not alter the temperature in the superburst ignition region. It may be interesting to search for transients that show superbursts during an accretion outburst because the quiescent cooling may potentially indicate the existence of extra heating and its depth. A comparison of the strength and depth of extra heating with the inferred superburst ignition depth in such sources would then allow constraints on the crust composition.

The nuclear properties of neutron-rich nuclei are important input for determining the strength of Urca cooling. Although the $Q_{\rm EC}$ values for nuclei in the ocean are determined from mass measurements to within $\approx 1\, \%$ \citep{audi2012}, $ft$-values and excited-state energies are more uncertain. Shell model calculations of all odd-$A$ nuclei in $rp$-process burning ashes are needed to determine excited-state energies and transition strengths to higher accuracy. Furthermore, experimentally determined $ft$-values and mass measurements of neutron-rich nuclei, for instance, at the Facility for Rare-Isotope Beams, will allow more accurate determinations of Urca pair neutrino luminosities and thereby improve constraints on the interior of superbursting neutron stars.

\acknowledgments
Support for A.D. and E.F.B. was provided by the National Aeronautics
and Space Administration through Chandra Award Number TM5-16003X
issued by the \emph{Chandra X-ray Observatory} Center, which is operated by
the Smithsonian Astrophysical Observatory for and on behalf of the
National Aeronautics and Space Administration under contract
NAS8-03060. A.D. and E.F.B. are also supported by the National Science Foundation under Grant No. AST-1516969. A.D. is also grateful for the support of the Michigan State University College of Natural Science Dissertation Completion Fellowship. Z.M. is supported by the National Science Foundation under Grant No. PHY-1419765 (Nuclear Structure and Nuclear Astrophysics). H.S. is supported by the US National Science Foundation under Grant No. PHY 11-02511. A.C. is supported by an NSERC Discovery grant and is a member of the Centre de Recherche en Astrophysique du Qu\'ebec
(CRAQ) and an Associate of the CIFAR Cosmology and Gravity program.
The authors are grateful for support received as part of the
International Team on Nuclear
Reactions in Superdense Matter by the International Space Science Institute in Bern, Switzerland. This material is based on work supported by the
National Science Foundation under Grant No. PHY-1430152 (Joint Institute for Nuclear Astrophysics \textrm{--} Center for the Evolution of the Elements).

\bibliographystyle{apj}

\end{document}